\documentclass[twocolumn,prb,epsfig,aps]{revtex4-1}
\usepackage{graphicx}
\usepackage{color}

\newcommand{\beq}{\begin{eqnarray}}
\newcommand{\eeq}{\end{eqnarray}}

\begin{document}
%\date{\today}
\setcounter{page}{1}
\title{Fluid and registered phases in the second layer of $^{\bf 3}$He on graphite}
\author{M.C. Gordillo}
\affiliation{Departamento de Sistemas F\'{\i}sicos, Qu\'{\i}micos y Naturales,
Universidad Pablo de Olavide. E-41013 Seville, Spain}
\author{J. Boronat}
\affiliation{Departament de F\'{\i}sica, 
Universitat Polit\`ecnica de Catalunya, 
Campus Nord B4-B5, E-08034 Barcelona, Spain}

\begin{abstract}
A quantum Monte Carlo approach, considering all the corrugation 
effects, was used to calculate the complete phase diagram of the second $^3$He 
layer adsorbed on graphite.  We found that  
a first-layer triangular solid was in equilibrium with  
a gas in the second layer. 
At a surface density $0.166 \pm 0.001$ \AA$^{-2}$, this 
fluid changes into two first-layer registered phases:   4/7  and 
7/12 solids. The 7/12 arrangement transforms into an  
incommensurate triangular structure of $\rho = 0.189 \pm 0.001$ \AA$^{-2}$ upon further helium loading. A 
recently proposed hexatic phase was found to be unstable with 
respect to those commensurate solids. 
\end{abstract}

%\pacs{05.30.Jp, 67.85.-d}

\maketitle
%\section{Introduction}

Helium adsorbed on graphite at temperatures close to zero is a standard setup to study 
the properties of stable quasi-two-dimensional quantum fluids and 
solids~\cite{cole}. The interplay between the additional third dimension, the effects
of the corrugation of the different substrates, and the quantum statistics of the adsorbed isotope 
($^4$He is a boson and $^3$He a fermion) produce  
very rich phase diagrams.  
In particular, 
there is a wealth of experimental data on the behavior of $^3$He atoms adsorbed 
on graphite, both on the first 
(clean or preplated) and second layers 
\cite{vilches1,greywall, greywall2,godfrin,godfrin2,saunders,godfrin3,godfrin4,godfrin5,fukuyama3,fukuyama1,smart,fukuyama,fukuyama4}. 
Recent experimental~\cite{smart,fukuyama} and theoretical~\cite{bonninhe3,yo4} 
work suggests that the low-temperature phase diagram of the first 
layer includes a liquid-gas coexistence, followed by a solidification at 
high $^3$He densities (first to a $\sqrt 3 \times \sqrt 3$ registered solid and 
then to 
an incommensurate one via another set of commensurate structures), and by 
second-layer promotion \cite{greywall}. It is worth noticing that the 
transition from gas to liquid is rather unique in quantum fluids and 
can be properly modeled only if the corrugation of the substrate is fully taken into account.  
Once promoted to the second layer, $^3$He atoms appear to be in a 
fluid-like phase, that eventually turns into a solid upon increase in the amount 
of helium adsorbed~\cite{greywall,godfrin2,godfrin3,
saunders,godfrin5,fukuyama3,fukuyama4}. 

The theoretical knowledge of that second-$^3$He layer is limited to 
$^4$He-preplated graphite~\cite{bonninhe3,yo5}.
A full calculation of that phase diagram~\cite{yo5} produced a set of results that 
compared favorably with the experimental data, predicting the existence of a 
very dilute liquid
that, at higher densities, is in equilibrium with a registered 7/12 solid that 
progresses to the formation of an incommensurate triangular phase close to third 
layer promotion. The density of that commensurate structure compares very favorable to the  
one found experimentally
\cite{godfrin2,saunders,godfrin3,godfrin4,godfrin5,fukuyama3,che}.  
However, some caution has to be exercised in comparing the areal densities in a theoretical 
calculation (simply the number or atoms divided by the surface) to the same experimental 
magnitude.  
Calorimetric measurements are typically given in terms of coverage, i.e., as a certain 
amount in excess of the density corresponding to the first-layer commensurate 
$\sqrt 3  \times \sqrt 3$ solid. This could produce a sizable discrepancy (up to 8.5\% \cite{lauter}) with its 
neutron scattering counterparts, and therefore some uncertainties
in the comparison to our simulations, further complicated with the presence of defects that vary from a sample to another. 
On another quarter, those theoretical results also indicate that an accurate description of the phase diagram 
demands the consideration of both the corrugation of the substrates and the 
relaxation of the first-layer atoms from their crystallographic positions.

In this Rapid Communication, we will be concerned with the quantum Monte Carlo  
description of a $^3$He layer on top of an incommensurate $^3$He solid adsorbed 
on graphite. As in Ref. \onlinecite{yo5}, the description of 
the system will be as realistic as possible, including corrugation and 
relaxation effects. In addition, we also analyzed different first-layer 
densities to allow for a compression upon helium loading.
A recent suggestion~\cite{fukuyama4} about the observation of a stable hexatic 
$^3$He phase will be also considered.

%\section{Method}

The starting point of our microscopic approach is the Hamiltonian for a system 
with two $^3$He layers adsorbed on  graphite,  
\begin{equation} \label{hamiltonian}
H = -\frac{\hbar^2}{2m} \sum_{i=1}^{N}  \nabla_i^2  
+ \sum_{i=1}^N V_{\text{ext}}({\bf r}_i)   
+ \sum_{i<j}^N V (r_{ij}) \ ,
\end{equation}
where $m$ is the $^3$He mass, and $N$ represents the total number of atoms 
(on both first and second layers)  at positions ${\bf r}_i$. The 
second term in Eq. (\ref{hamiltonian}) corresponds to the sum of all individual 
C-He interactions, modeled
by the accurate Carlos and Cole interatomic potential~\cite{carlosandcole}.   
The graphite sheets containing the carbon atoms were simulated in the same way 
as  
in previous literature~\cite{yo4,yo5,yo,yo2,yo3,jltp}. $V(r_{ij})$ stands for 
the helium-helium Aziz potential~\cite{aziz}, a standard of the theoretical 
descriptions of helium at low temperatures. Here, $r_{ij}$ is the distance 
between any two helium atoms, 
irrespectively of their location on the first or second layer.  

We solved the many-body Schr\"odinger equation associated to the Hamiltonian
of Eq. (\ref{hamiltonian}) by using the fixed-node Diffusion Monte Carlo 
(FN-DMC) method~\cite{hammond}. This technique provides us with an approximation 
to the ground state of the system, ground state that it is expected to be a good
description of the real experimental setup at the mK temperatures characteristic 
of these studies. 
The sign problem of a Fermi system like a set of $^3$He atoms prevents an exact calculation, 
in opposition to what happens for bosonic $^4$He. However, the fixed-node approximation 
is a stable technique that furnishes us with an upper bound for the ground-state energy 
of a system of fermions. In the FN-DMC method, the nodes of the ground-state wavefunction 
are imposed to be the same as the ones of a trial wavefunction (initial approximation to the real 
wavefunction). %This means that if we know 
%the nodal surface of the exact wavefunction, the exact energy is readily accessible.  
Unfortunately, the 
position of the real nodes is unknown {\em a priori}, but the use of an 
accurate trial function could leave that upperbound very close to the 
real value.    
We used %as trial wave function
\begin{eqnarray}
\Phi({\bf r}_1, {\bf r}_2, \ldots, {\bf r}_N) = \Psi_u({\bf r}_1, {\bf
r}_2, \ldots, {\bf r}_{N_u}) \times \nonumber  \\
\Psi_d({\bf r}_{N_u + 1}, {\bf r}_{N_u+2}, \ldots, {\bf r}_{N}),
\label{trialtot}
\end{eqnarray}
with ${\bf r}_1, {\bf r}_2, \ldots, {\bf r}_{N_d}$ 
the coordinates of the $N_u$ helium atoms on the second layer, and 
${\bf r}_{N_u + 1}, {\bf r}_{N_u+2}, \ldots, {\bf r}_{N}$ the ones for the $N_d= 
N-N_u$ atoms in direct contact with graphite.  
Following Ref. \onlinecite{yo5}, we considered as trial wave function for the 
upper layer    
\begin{eqnarray}
\Psi_u({\bf r}_1, {\bf r}_2, \ldots, {\bf r}_{N_u}) =  
D^{\uparrow} D^{\downarrow}  \prod_i^{N_u}  u_u({\bf r}_i)  \times \nonumber \\
\prod_{i<j}^{N_u} \exp \left[-\frac{1}{2} 
\left(\frac{b}{r_{ij}} \right)^5 \right] \ , 
\label{trialu}
\end{eqnarray}
with  $D^{\uparrow}$ and $D^{\downarrow}$  the  two-dimensional Slater
determinants for spin up and down atoms, respectively.
The coordinates of the particles included in those determinants were corrected
by backflow terms in the standard way,
\begin{eqnarray}
\tilde x_i  & = & x_i + \lambda \sum_{j \ne i} \exp [-(r_{ij} -
r_b)^2/\omega^2] (x_i - x_j) \\
\tilde y_i  & = & y_i + \lambda \sum_{j \ne i} \exp [-(r_{ij} - r_b)^2/\omega^2] (y_i - y_j).
\end{eqnarray} 
Here,  
$\lambda = 0.35$; $\omega =
1.38$ \AA,  and $r_b = 1.89$ \AA \cite{yo4,casulleras}.   
We considered unpolarized systems, i.e., $N_d = N_u = 
N/2$. The function $u_u({\bf r})$ is the numerical 
solution of the one-body 
Schr\"odinger equation that describes a single $^3$He atom on top of a
triangular lattice formed by first-layer $^3$He atoms located in the crystallographic
positions of an 
incommensurate triangular phase, neglecting the influence of the graphite 
structure~\cite{yo4}. 
In the present
work, we used three first layer triangular lattices of  
densities taken from different experimental works:  0.109 (from Ref. \onlinecite{greywall}), 0.113 (intermediate from those of Ref. \onlinecite{lauter} and 
\onlinecite{ziouzia}) and 0.116 \AA$^{-2}$ (from Ref. \onlinecite{fukuyama4}). This was done in order to take into account a possible
compression of the bottom layer upon increasing of the overall helium density.  
The value of $b$ was optimized variationally ($b$ = 2.96 \AA).

The bottom-layer trial wave function was 
\begin{eqnarray}
\Psi_d({\bf r}_{N_u + 1}, {\bf r}_{N_u+2}, \ldots, {\bf r}_{N})  = 
\prod_i^{N_d}  u_d({\bf r}_i) \times  \nonumber \\ 
\prod_{i<j}^{N_d} \exp \left[-\frac{1}{2}
\left(\frac{b}{r_{ij}} \right)^5 \right]   \times \nonumber  \\
\prod_i^{N_d} \exp \left\{ -a [(x_i-x_{\rm site})^2 + (y_i-y_{\rm site})^2]
\right\}  
\label{triald}
\end{eqnarray}
The last (Nosanov) term compels the atoms to stay close around 
their crystallographic positions $\{x_{\rm site},y_{\rm site} \}$. 
$u_d({\bf r}_i)$ is the
numerical solution of the one-body  Schr\"odinger equation for one $^3$He
atom on top of graphite, and $a$ was set to 0.24 \AA$^{-2}$ as in previous 
work~\cite{yo4}.  
The solid phases of the upper layer were also simulated by multiplying 
Eq. (\ref{trialu})   by a Nosanov term, as in Eq. (\ref{triald}).

The possible existence of an hexatic phase in the second layer was also 
studied by multiplying Eq. (\ref{trialu}) by \cite{apaja} 
\begin{equation}
\psi_h= \prod_{i<j}^{N_u} \exp \left[ \alpha \frac{\cos(m \phi_{ij}) -1}{r_{ij}} 
\right],
\end{equation}  
with $\cos(\phi_{ij}) = \frac{{\bf r_j} - {\bf r_i}}{r_{ij}}$.  
The value of the variational constants
$m$ and $\alpha$ was taken from Ref. \onlinecite{apaja}.    

%\section{Results} 

\begin{figure}
\begin{center}
\includegraphics[width=0.8\linewidth]{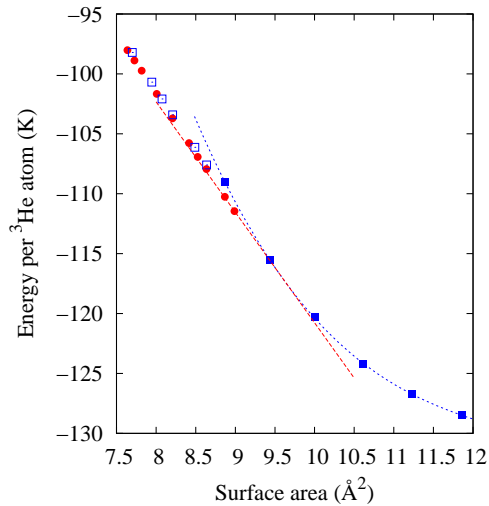} 
\caption{(Color online) Energy per $^3$He atom  
as a function of the inverse of the total $^3$He density. Full squares, 
a single layer incommensurate solid; full circles, a second layer system
including a first layer solid of density 
0.109 \AA$^{-2}$;  open squares, second layer arrangement on top 
of a 0.113 \AA$^{-2}$ incommensurate triangular phase.  
Dotted line, third order polynomial fit to the single layer energy values, 
intended as a
guide-to-the-eye; dashed line, double-tangent Maxwell line 
to determine the coexistence between phases.
}

\label{fig1}
\end{center}
\end{figure}

We followed here the methodology used previously to describe two $^4$He layers
adsorbed on graphene~\cite{doble}. Thus, we started by considering different 
first-layer triangular lattice densities and 
calculated the energies for the entire set of atoms (irrespectively of their location on the first or second layer). The
first set of data, corresponding to low second-layer densities, is displayed in 
Fig.~\ref{fig1}. In order to 
obtain the stability ranges of the different phases, we performed double-tangent Maxwell constructions using our
FN-DMC results. This means that the $x$-axis in Fig.~\ref{fig1} represents 
the inverse of the total 
(first + second layer) density. In that figure,   
full squares were taken from Ref. \onlinecite{yo4} and correspond 
to the incommensurate solid phase of the (single) first layer.
Full circles 
stand by the results from a simulation including $16 \times 9$ triangular 
lattice cells ($52.08 \times 50.74$ \AA$^2$; 288 $^3$He atoms) 
in the first layer plus the necessary atoms in the second one to account for the displayed surface per atom. This corresponds  to  
a bottom layer of density $\rho=0.109$ \AA$^{-2}$, in line with 
experimental results of Ref. \onlinecite{greywall}.
Open squares are simulation data for a first layer comprising $14 \times 8$ 
similar cells to give us a bottom layer density of 0.113 \AA$^{-2}$
($44.8 \times 44.34$ \AA$^2$; 224 atoms).
It can be seen that the open squares are consistently above the open circles in the inverse density range displayed.  
Therefore, we should draw the double-tangent Maxwell line (dashed line in 
Fig.~\ref{fig1}) between the 0.109 \AA$^{-2}$ data and the results for a single 
layer solid. From that line, we can establish 
that a single-layer structure of density 0.106 $\pm$ 0.002 \AA$^{-2}$ is in equilibrium with 
a two-layer system with total density 0.111 $\pm$ 0.002 \AA$^{-2}$. This means that from 0.106 \AA$^{-2}$
up,
one would have a mixture of clean first-layer zones with very dilute second layer
systems of 0.111-0.109 = 0.002 \AA$^{-2}$ in the adequate proportions to produce total 
intermediate densities in the range from 0.106 \AA$^{-2}$ to 0.111 \AA$^{-2}$. 
This is in excellent agreement with the 
experimental values given in Refs. \onlinecite{vilches1} ($\sim$ 0.108 \AA$^{-2}$), \onlinecite{godfrin3} 
($\sim$ 0.106 \AA$^{-2}$) and \onlinecite{lauter} ($\sim$ 0.105 \AA$^{-2}$), obtained with different techniques.

\begin{figure}
\begin{center}
\includegraphics[width=0.8\linewidth]{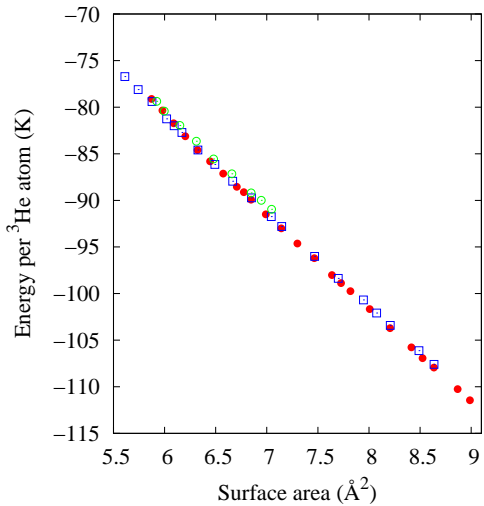} 
\caption{(Color online) 
Energy per $^3$He atom for three second-layer arrangements, with different 
densities of the
underlying first-layer solid, as 
a function of the inverse of the total density.   
Full circles, $\rho = 0.109$ \AA$^{-2}$;  
open squares, $\rho =  0.113$ \AA$^{-2}$; open circles, $\rho = 0.116$ 
\AA$^{-2}$. 
}
\label{fig2}
\end{center}
\end{figure}

The energy when we increase the density (or decrease the surface per atom), 
is displayed in Fig.~\ref{fig2}. The symbols are 
the same as in Fig.~\ref{fig1}, but we included a third 
set of calculations (open circles) in which the underlying incommensurate 
solid density 
was 0.116 \AA$^{-2}$, following the experimental findings of Ref. 
\onlinecite{fukuyama4}. Two things are immediately apparent: first, this last 
setup is always 
metastable with respect to the first two arrangements, and second, on increasing 
the helium density, the energies corresponding to the open squares 
start to go below the ones represented by full circles. This means that the 
first layer solid undergoes a compression upon helium loading.
This is in line with previous results for a double $^4$He layer on graphene. 
Third-order polynomial fits to the data in Fig.~\ref{fig2}, not shown for 
simplicity, indicate that the crossing is produced at a density $\rho=0.156 
\pm 0.002$ \AA$^{-2}$ 
($6.41 \pm 0.01$ \AA$^2$ in Fig.~\ref{fig2}). This corresponds to a 
second-layer density of $0.045 \pm 0.002$ \AA$^{-2}$.

\begin{figure}
\begin{center}
\includegraphics[width=0.8\linewidth]{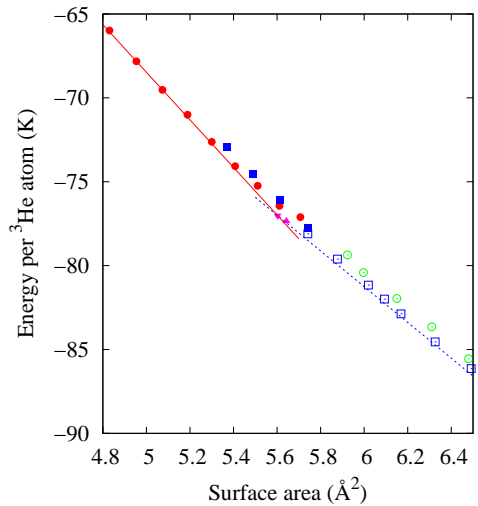} 
\caption{(Color online)
Same as in the previous figure, but including a double-layer incommensurate 
solid
(upper full circles), an hexatic phase (full squares), a $4/7$ (up triangle) and 
a $7/12$ (down triangle) 
commensurate phases on top of a first-layer solid of density 0.113 \AA$^{-2}$.  
Full and dotted lines correspond to Maxwell constructions between the different 
stable phases. 
}
\label{fig3}
\end{center}
\end{figure}

Fig.~\ref{fig3} reports the coexistence between the second-layer fluid (open 
squares), and the 4/7 and 7/12 registered structures with the first layer
(full triangles). The dotted line is a double-tangent Maxwell line between the 
4/7 solid and a fluid of density $0.166 \pm 0.001$ \AA$^{-2}$
($6.01 \pm 0.01$ \AA$^2$ in Fig.~\ref{fig3}). In both cases, the underlying 
first-layer density was 0.113 \AA$^{-2}$, since a more compressed
triangular solid increases the overall energy per particle. This result is in good
agreement with the experimental 0.111 \AA$^{-2}$ value provided in Ref. \onlinecite{lauter}. 
Since the line can be prolonged to higher densities to include the 7/12 structure, we conclude that our 
results support the coexistence between {\em both} registered solids and the 
0.166 \AA$^{-2}$ fluid. This is not surprising given their very close densities ($\rho_{4/7} = 
0.177$ \AA$^{-2}$, 
$\rho_{7/12} = 0.179$ \AA$^{-2}$). In the same figure, full circles stand for the 
energy results for the hexatic phase recently proposed in Ref. 
\onlinecite{fukuyama4} to account for the experimental data.     
As we can see, those data are above both the results of the commensurate structures, and the second-layer incommensurate triangular solid represented by the full circles. 
This means that this phase is unstable with respect to any of those solids, at 
least in the limit  $T= 0$.

The full
line in Fig.~\ref{fig3} is another double-tangent Maxwell line, this time 
between the 7/12 structure and a second-layer incommensurate solid of $\rho = 
0.189 \pm 0.001$ \AA$^{-2}$ ($5.30 \pm 0.01$ \AA$^2$ in Fig.~\ref{fig3}).
This density is rather close to the one corresponding to the third-layer 
promotion observed in different experiments ($\rho = 0.184$ \AA$^{-2}$ 
Ref. \onlinecite{greywall}; $\rho = 0.186$ \AA$^{-2}$, Ref. \onlinecite{saunders}; $\rho = 0.187$ \AA$^{-2}$, Ref. \onlinecite{siqueira}; 
$\rho = 0.19$ \AA$^{-2}$, Ref. \onlinecite{fukuyama3}).
Those values are smaller or compatible with the one deduced from our data in 
Fig.~\ref{fig3}. This implies that experimentally we should have an equilibrium between a 
clean second-layer 7/12 structure of 
$\rho = 0.179$ \AA$^{-2}$ and a setup with a low-density fluid on top of a second 
layer solid 
~\cite{greywall,saunders,fukuyama3}. 
The nature of 
the transformations undergone by the second layer upon further helium loading 
is beyond the scope of the present work.

\begin{figure}
\begin{center}
\includegraphics[width=0.8\linewidth]{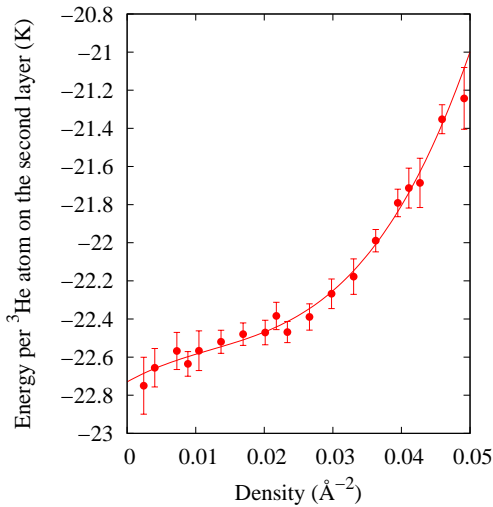} 
\caption{(Color online)
Energy per $^3$He atom on the second layer as a function of that layer density. 
The density of the first layer was $\rho = 0.109$ \AA$^{-2}$. 
Full line, third-order polynomial fit to the simulation data, intended 
exclusively as a guide-to-the-eye. 
}
\label{fig4}
\end{center}
\end{figure}

At this point, a remaining question is that of the nature of the
second-layer fluid before solidification. To solve that, we plotted
the energies per $^3$He atom {\em on the second layer} versus the $^3$He 
density on that layer alone. This is done in Fig.~\ref{fig4}. As one can see, 
our FN-DMC results correspond to a gas phase, since the energy 
per atom increases monotonically as a function of the $^3$He density, 
with no discernible plateau that would be the tell-tale signal of a liquid-gas 
transition~\cite{yo4}.

%\section{Discussion} 

In summary, we have undertaken the calculation of the rather complicated equation of state of the 
second layer of $^3$He on graphite. The comparison between 
previous theoretical descriptions of the same 
system on preplated graphite~\cite{bonninhe3,yo4} and the experimental data 
suggested  the necessity of including in the calculations all corrugation and 
dynamic effects.
That implies the consideration of different first-layer densities to take into
account a possible compression, in line with what happened for $^4$He on 
graphene. With all that, our quantum Monte Carlo 
results compare very favorably with the available experimental data. This is 
true for the second-layer promotion density $\rho = 0.106$ \AA$^{-2}$ 
~\cite{godfrin3}
and the upper density limit for a fluid  (0.053 \AA$^{-2}$ versus the 0.055 \AA$^{-2}$ of Ref. \onlinecite{greywall}, and the  0.050-0.060  \AA$^{-2}$ interval proposed
on Ref. \onlinecite{godfrin2}). The solidification into the registered 
structures is also well predicted ($\rho \sim 0.178$  \AA$^{-2}$, the same 
value that experiment~\cite{saunders,
fukuyama3}). This also validates our value for the first layer density upon
compression (0.113 \AA$^{-2}$), and differs from the one proposed in Ref. 
\onlinecite{fukuyama4} (0.116 \AA$^{-2}$). 

Our data also suggest that the registered 4/7 and 7/12 solids are in equilibrium 
with a three-layer system of $\rho \sim 0.19$ \AA$^{-2}$. This means that 
from $\rho = 0.179$ \AA$^{-2}$ up 
there is a mixture of a second-layer 4/7 and 7/12 structures and a third-layer fluid.  
Again this agrees with previous experimental findings 
\cite{godfrin5,fukuyama3}, but not with the suggestion
of a stable hexatic phase around the same density range. 
This means that the suggested hexatic phase can hardly be a 
candidate for the quantum spin liquid proposed in Ref. \onlinecite{fukuyama4}.

%\acknowledgments
We acknowledge partial financial support from the
Junta de Andaluc\'{\i}a group PAI-205 and and MINECO (Spain) Grants 
No. FIS2014-56257-C2-2-P, FIS2017-84114-C2-2-P, FIS2014-56257-C2-1-P, and 
FIS2017-84114-C2-1-P.

\end{document}